\documentclass[twocolumn,pre,showpacs,preprintnumbers,amsmath,amssymb,superscriptaddress]{revtex4}

\usepackage{graphicx}
\usepackage{dcolumn}
\usepackage{bm}
\usepackage{longtable}

\begin{document}

\preprint{APS/123-QED}

\title{L\'evy walks and scaling in quenched disordered media}

\author{}
\affiliation{}

\author{Raffaella Burioni}
\affiliation{Dipartimento di Fisica, Universit\`a degli Studi di
Parma, viale G.P.Usberti 7/A, 43100 Parma, Italy}
\affiliation{INFN, Gruppo Collegato di
Parma, viale G.P. Usberti 7/A, 43100 Parma, Italy}
\author{Luca Caniparoli}
\affiliation{International School for Advanced Studies SISSA, via Beirut 2/4, 
34151, Trieste, Italy}
\author{Alessandro Vezzani}
\affiliation{
Centro S3, CNR-Istituto di Nanoscienze, Via Campi 213A, 41125 Modena Italy}
\affiliation{Dipartimento di Fisica, Universit\`a degli Studi di
Parma, viale G.P.Usberti 7/A, 43100 Parma, Italy}

\date{\today}

\begin{abstract}
We study L\'evy walks in quenched disordered one-dimensional media, with scatterers spaced 
according to a long-tailed distribution. By analyzing the scaling relations for the random-walk probability 
and for the resistivity in the equivalent electric problem, we obtain the asymptotic behavior of the mean 
square displacement as a function of the exponent characterizing the scatterers distribution. 
We demonstrate that in quenched media different average procedures can display different asymptotic behavior. 
In particular, we estimate the moments of the displacement averaged over processes starting 
from  scattering sites. Our results are compared with numerical simulations, with 
excellent agreement.
\end{abstract}

 \maketitle

Random walks in quenched random environments occur in many fields of
statistical and condensed matter physics  \cite{bouchaud}, as  they represent
the simplest model of diffusion phenomena and non-deterministic motion.  
Disorder and geometrical confinement are known to strongly influence transport properties. In particular, in highly spatial inhomogeneous media, the diffusion process is often characterized by large distance diffusion events, which play a crucial role in transport phenomena and can strongly enhance them \cite{klages}. Molecular diffusion at low pressure in porous media is dominated by collision with pore walls, with ballistic motion inside the large pores \cite{knudsen}, diffusion in chemical space over polymer chains can be described by a distribution of step length with power law behavior \cite{sokolov}.
In addition, recent experiments on new disordered optical materials paved the way  to a tuned engineering of L\'evy-like distributed step lengths \cite{Levy}. These and many other processes can often be successfully analyzed using the  L\'evy walks formalism  \cite{shles1985}: The random walker can perform long steps, whose distribution is characterized by a power law 
behavior $\lambda(r)\sim 1/r^{\alpha+1}$, with $\alpha>0$, for large distance displacements $r$.  

An important feature of these experimental settings is that the random walk 
is in general correlated, and the correlation is induced by the topology of the quenched medium. 
Diffusing agents moving in highly inhomogeneous regions, where they just experienced a long distance jump without being scattered, have a high probability of being
backscattered at the subsequent step undergoing a jump
of similar size, and this leads to a correlation in step lengths. While the effect of annealed disorder on transport properties in  L\'evy walks is quite well understood \cite{ann}, the role of correlations in L\'evy-like motion is still an open problem.

If the motion occurs in low dimensional
samples, spatial correlations in jump probabilities can deeply influence  the 
diffusion properties. This was first evidenced in models of L\'evy flights, 
\cite{Fogedby94} and more recently discussed in one-dimensional 
models for  L\'evy walks on quenched and correlated random
environments. The recent studies 
focused, respectively, on the mean square displacement
in a L\'evy-Lorentz gas \cite{klafter}, and on the conductivity and 
transmission through a chain of barriers with L\'evy-distributed spacings 
\cite{beenakker}. Both studies evidenced the effect of correlations on large scale quantities:
the mean square displacement and the conductivity indeed feature a modified scaling behavior
with respect to the annealed case, and similar results have been obtained for a deterministic 
fractal characterized by holes distributed according to a L\'evy statistics \cite{noi}.

\begin{figure}
\includegraphics[width=\columnwidth]{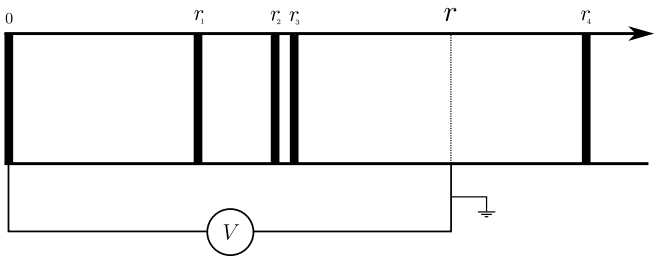}
\caption{The L\'evy walk model and the equivalent electric problem. Scatterers 
are placed at positions $o,r_1,r_2\dots$.  The spacings $r_ {j+1} - r_ {j}$
between scatterers are L\'evy distributed according to the probability density (\ref{eq:p_r}).}
\label{model}
\end{figure}

We consider here an analogous model. In particular we introduce 
a one-dimensional structure where we place scatterers spaced according to a 
L\'evy-type distribution (Figure \ref{model}) so that the 
probability density for two consecutive scatterer to be at distance $r$ is
\begin{equation}
  \label{eq:p_r}
  \lambda(r)\equiv \alpha r_0^\alpha \frac{1}{r^{\alpha+1}}, \quad r\in [r_0,\infty),
\end{equation}
where $\alpha>0$ and $r_0$ is a cutoff fixing the scale-length of the system.
A continuous time random walk \cite{ctrw} 
is naturally defined on this structure, i.e.
a walker moves ballistically (at constant velocity $v$) until it reaches one of 
the scatterers, and then it is transmitted or reflected with probability $1/2$.
Figure \ref{model}  also describes the equivalent electrical model, where the 
resistance between  two points  is defined as
the number of scatterers between them \cite{beenakker}. 

A subtle effect of quenched disorder is the dependence of  the observables on
the choice of the walker starting site or, equivalently,  of the position of 
the electric contacts. In particular, on inhomogeneous structures, averages 
over different starting points can provide different results  
with respect to the corresponding local 
quantities \cite{burioni,noi}. 
In presence of  long tails, local quantities are strongly fluctuating and 
averages are in some sense necessary 
for a global characterization of the structure. 
In our system, two different average procedures can be 
introduced and interestingly, due to  L\'evy statistics, 
they provide different results \cite{klafter}.

The first possibility is to consider as a starting site at $t=0$ any point 
of the structure and then average over all stochastic processes
and realizations of the disorder. In this 
situation the behaviour is quite clear \cite{klafter}. For $\alpha<1$ the motion is 
always ballistic, while for $\alpha>1$ it can be shown that the probability of
reaching a site at distance $r$ at the first step decays as $1 / r^{\alpha}$ (i.e. 
much more slowly than $\lambda(r)$) \cite{klafter}; hence, the motion can be dominated by the first 
jump (called ballistic peak). In particular, for $1\leq \alpha \leq 2$ the 
first step provides the major contribution to the mean square displacement, i.e.
$\langle r^2 (t) \rangle \sim C t^{3-\alpha}$.

In experimental realizations however, processes begin with a scattering event,
corresponding to the entrance of particles or light in the system \cite{Levy} 
and, therefore, averages should be performed considering only scatterers as 
starting points.  In this case the static behaviour of the resistance has 
been calculated in \cite{beenakker}, while the situation in the study of the 
dynamics of random walks is much more obscure. Indeed, the probability 
of reaching a distance $r$ at the first step is the same as that of 
any other scattering event. Therefore, the first-jump does not determine
the behaviour of the means square displacement and only a lower bound
can be proved 
\cite{klafter}: $\langle r^2 (t) \rangle > v^2t^2P^+(t)\sim t^{2-\alpha}$, 
where $P^+(t)$ is the probability for the first step to be longer than $t$. 
An insight on the issue would represent a relevant result 
both for a theoretical understanding of quenched L\'evy processes both for guiding future 
experimental investigations.

In this paper, using the analytic result for the resistance \cite{beenakker},
the mapping between resistance and random walks  \cite{doylesnell}, a  
scaling hypothesis for the probability distribution $P(r,t)$ \cite{cates}
and the reasonable hypothesis of ``single long jump'', we derive an analytic 
expression for the asymptotic behaviors of the characteristic length $\ell (t)$
of $P(r,t)$ and for the mean square displacement $\langle r^2 (t) \rangle$, 
when averaged over the scattering points. The predicted behaviors are
\begin{equation}
\ell (t) \sim
\begin{cases}
t^{\frac{1}{1+\alpha}} & \mathrm{if}\ 0<\alpha<1 \\
t^{1 \over 2} & \mathrm{if}\ 1 \leq \alpha  
\label{ellcom}
\end{cases}
\end{equation}
and
\begin{equation}
\langle r^2 (t) \rangle \sim
\begin{cases}
t^{\frac{2+2 \alpha-\alpha^2}{1+\alpha}} & \mathrm{if}\ 0<\alpha<1 \\
t^{\frac{5}{2}- \alpha} & \mathrm{if}\ 1 \leq \alpha \leq 3/2 \\
t & \mathrm{if}\ 3/2 < \alpha  
\label{r2com}
\end{cases}
\end{equation}
Our predictions are compared with extensive numerical simulations, 
with excellent agreement. 
We note that for $0< \alpha< 3/2$  the system does not follow
the standard behaviour $\langle r^2 (t) \rangle \sim \ell(t)^2$. 

Let us introduce our argument. In a one dimensional system the most general 
scaling hypothesis for $P(r,t)$ is:
\begin{equation}
P(r,t)=\ell^{-1}(t)f(r/\ell(t))+g(r,t) 
\label{sal}
\end{equation}
with  a convergence in probability
\begin{equation}
\lim_{t\to\infty}\int_0^{v t} |P(r,t)-\ell^{-1}(t)f(r/\ell(t))|dr=0
\label{sal2}
\end{equation}
The integration cut off is provided by the fact that the walker in a 
time $t$ covers at most a distance $vt$ ($v$ is the velocity).
The leading contribution to $P(r,t)$  is hence  
$\ell^{-1}(t)f(r/\ell(t))$ which is significantly different from 
zero only for $r \lesssim \ell(t)$. 
The subleading term $g(r,t)$, with the condition $\lim_{t \to \infty}\int |g(r,t)| dr=0$, describes the behavior 
at larger distances, i.e. $\ell(t)\ll r < t$. Notice that, 
if $g(r,t)$ does not vanish rapidly enough, 
it can nevertheless provide important contributions to $\langle r^2(t)\rangle$.

The scaling of the correlation length $\ell(t)$ is given by:
\begin{equation}
\ell(t)\sim t^{d_s/2}
\label{sal3}
\end{equation}
where we denoted  the exponent as $d_s/2$, in order to obtain $P(0,t)\sim t^{-d_s/2}$,
i.e. the standard behavior of the return probability for random walks on graphs \cite{Orbach}.
In \cite{noi,cates} it is shown that, in the equivalent electric network problem,
the resistance $R(r)$ between sites at  distance $r$ corresponds to the random walk
quantity $\lim_{\omega \to 0}\tilde P(r,\omega)$ where $\tilde P(r,\omega)$ is the time 
Fourier transform of $P(r,t)-P(0,t)$. 

Since $g(r,t)$ is relevant only in the regime
$r\gg\ell(t)$ and the resistance is evaluated 
at a finite distance $r$ and for a diverging characteristic 
length $\ell(t)$ (i.e. $\omega \to 0$), it is clear that
$g(r,t)$ does not provide significant contributions to $R(r)$.
Therefore, analogously to \cite{noi,cates}, the scaling of the resistance is given by
\begin{equation}
R(r)\sim r^{2/d_s-1}.
\label{scalR}
\end{equation}
The stationary problem and the calculus of the resistance are a much more simple 
task than the direct dynamical study of random walks, so that the asymptotic 
behavior of $R$ at large distances can be calculated analytically
\cite{beenakker} obtaining
$R(r)\sim r^{\alpha}$ for $\alpha<1$ and $R(r)\sim r$ for $\alpha>1$. 
Plugging this result into equations (\ref{sal3}-\ref{scalR}) we obtain
the asymptotic behaviour of $\ell(t)$ (\ref{ellcom}).

The explicit expression of the mean square displacement reads
\begin{equation}
\langle r^2(t)\rangle=\int_0^{vt} \ell^{-1}(t) f(r/\ell(t))r^2dr
+\int_0^{vt} g(r,t) r^2 dr.
\label{r2}
\end{equation}
Anomalies with respect to the usual behavior 
$\langle r^2(t) \rangle \sim \ell^2(t)$, can be present for two reasons.
First, the second term in (\ref{r2}) can be dominant with respect the first one 
(this happen e.g. when averaging over any starting site \cite{klafter}). 
A second more subtle anomaly shows up if the scaling function
$f(x)$ decays too slowly for $x\to\infty$, as in the case of the 
annealed model \cite{ann}. We will show that, depending on the value of $\alpha$, 
our system presents both situations.

Since the anomalies are determined by the regime where $r\gg \ell(t)$, 
we expect that the dynamics is dominated by a single long jump of length $r$.
In averages over all starting sites the probability of long jumps 
is much higher at the first step, hence the single long range event occurs 
at $t=0$. 
Here on the other hand, it can happen, with equal probability, at any 
scattering site. 
In particular, neglecting the possibility of multiple ``long jumps'' we obtain,
for $r\gg \ell(t)$, $P(r,t) \sim N(t)/r^{1+\alpha}\ll 1$, 
where $N(t)$ is the number
of scattering sites visited by the walker in a time $t$ and $1/r^{1+\alpha}$
is the probability for a scatterer to be followed by a jump of length $r$. 
Discarding the single long jump, the distance crossed by the walker in 
time $t$ is of order $\ell(t)$. According to the results in \cite{beenakker}, 
the number of scattering sites visited in this time is the 
resistance of a segment of length $\ell(t)$, i.e.  $\ell(t)^\alpha$ for
$\alpha<1$ and $\ell(t)$ for $\alpha \geq 1$. 
This implies that $N(t)\sim t^{\alpha/(1+\alpha)}$ for
$\alpha<1$ and $N(t)\sim t^{1/2}$ for $\alpha\geq 1$. 

For $\alpha<1$ and $r\gg \ell(t)$, we estimate  $P(r,t)$ as
\begin{equation} 
P(r,t) \sim t^{\alpha \over {1+\alpha}}\frac{1}{r^{1+\alpha}} \sim
\frac 1 {\ell(t)} \left( \frac {r} {\ell(t)}\right)^{-(1+\alpha)}
\label{Plmin1}
\end{equation}
Hence the scaling function $f(x)$ features a long tail for large $x$ decaying
as $x^{-1-\alpha}$. On the other hand for $\alpha\geq1$ and $r\gg \ell(t)$ 
we obtain
\begin{equation}
P(r,t) \sim t^{1 \over 2} \frac{1}{r^{1+\alpha}} \sim
\frac 1 {t^{(\alpha-1)/2} \ell(t)} \left( \frac {r} {\ell(t)}\right)^{-(1+\alpha)} \sim g(r,t)
\label{Plmag1}
\end{equation}
Note that here $g(r,t)$ provides a subleading contribution to $P(r,t)$. Indeed 
$\lim_{t \to \infty} \int_{\ell(t)}^{vt}  g(r,t)=0$ ($\alpha>1$).

\begin{figure}
\includegraphics[width=0.5\textwidth]{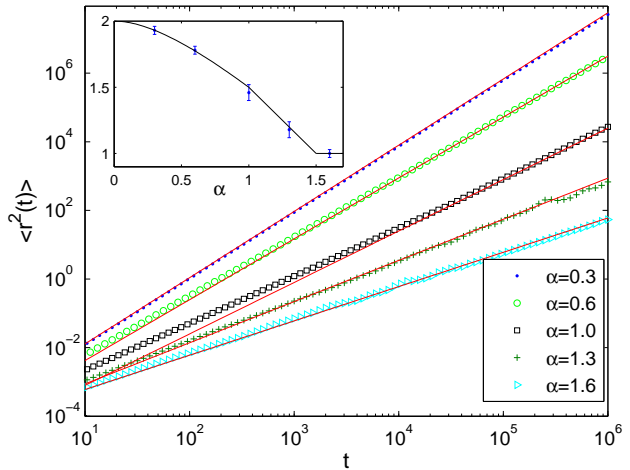}
\caption{(Color on-line) Mean-square displacements as a function of time for different values of $\alpha$, i.e. symbols. Theoretical predictions given by equations (\ref{r2com}) are represented by thin (red) lines. Theoretical values of the exponent as a function of $\alpha$ is compared with the fitted values in the inset.}
\label{r2f}
\end{figure}

\begin{figure}
\includegraphics[width=0.5\textwidth]{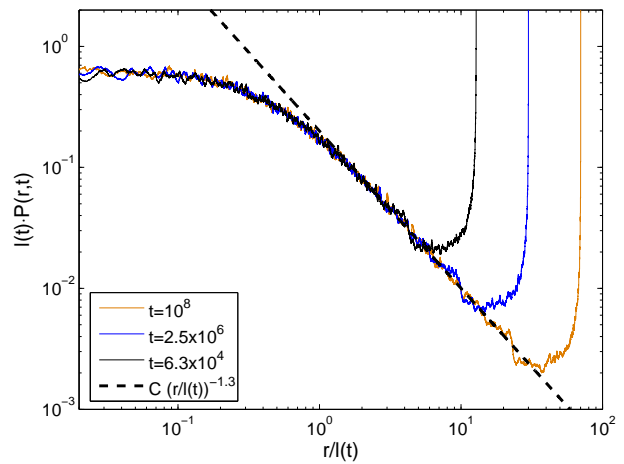}
\caption{(Color on-line) Montecarlo evaluation of the probability density rescaled by $\ell(t)$ as a function of $r/\ell(t)$ for $\alpha=0.3$. The dashed black 
line evidence that for $r>\ell(t)$ $P(r,t)$ features the behavior described in  (\ref{Plmin1}). Spikes correspond to ballistic peaks at $r=t$, which provides the subleading contribution $\langle r^2\rangle \sim t^{2-\alpha}$ to the mean square displacement.}
\label{Prt0_3}
\end{figure}

\begin{figure}
\includegraphics[width=0.5\textwidth]{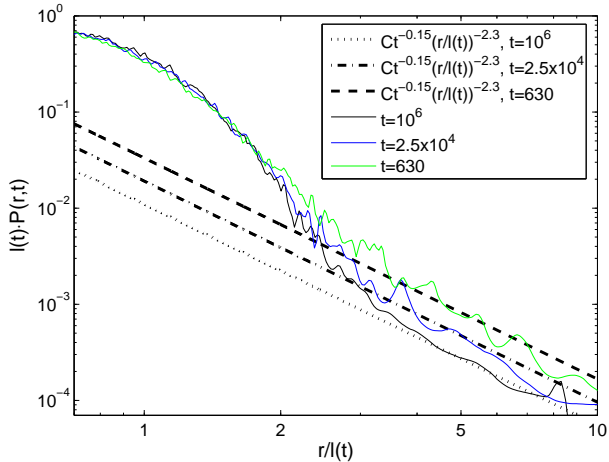}
\caption{(Color on-line) Montecarlo evaluation of the probability density rescaled by $\ell(t)$ as a function of $r/\ell(t)$ for $\alpha=1.3$. The black 
lines evidence that for $r>\ell(t)$ $P(r,t)$ features the behavior described in  (\ref{Plmag1}).}
\label{Prt1_3}
\end{figure}

We note that the contribution  to $\langle r^2(t)\rangle$ of 
lengths $r \lesssim \ell(t)$ is always of order $\ell(t)^2$,
while, at larger distances, the dominant contribution is provided 
by probabilities  (\ref{Plmin1}-\ref{Plmag1}). The contributions coming from
these tails are, for $\alpha<1$
\begin{equation} 
\int_{\ell(t)}^{v t} t^{\alpha \over {1+\alpha}}\frac{r^2}{r^{1+\alpha}} dr \sim
t^{\frac{2+2 \alpha-\alpha^2}{1+\alpha}}
\label{r2min1}
\end{equation}
and for $\alpha>1$
\begin{equation}
 \int_{\ell(t)}^{vt} t^{1 \over 2} \frac{r^2}{r^{1+\alpha}} dr \sim
t^{\frac{5}{2}-\alpha}.
\label{r2mag1}
\end{equation}
The contribution  (\ref{r2min1}) for large times  is always greater than 
$\ell^2(t)$ , while (\ref{r2mag1}) is dominant with respect $\ell^2(t)$ only for $\alpha<3/2$. 
The overall behavior of the mean square displacement is therefore
given by (\ref{r2com}).

Analogously we estimate the moments 
$\langle r^p(t)\rangle$, $p>0$,
\begin{equation}
\langle r^p (t) \rangle \sim
\begin{cases}
t^{\frac{p}{1+\alpha}}\sim \ell(t)^p & \mathrm{if}\ \alpha<1,
\ p<\alpha \\
t^{\frac{p(1+\alpha)-\alpha^2}{1+\alpha}} & \mathrm{if}\ \alpha<1, 
\ p>\alpha \\
t^{\frac{p}{2}}\sim \ell(t)^p & \mathrm{if}\ \alpha>1,
\ p<2 \alpha-1 \\
t^{\frac{1}{2}+p-\alpha} & \mathrm{if}\  \alpha>1,
\ p>2 \alpha-1 
\label{rpc2}
\end{cases}
\end{equation}

We now compare our finding with extensive numerical simulations.
We set the cutoff for  the length scale of the system,  $r_0=0.1$ and
the velocity in the ballistic stretches $v=1$.
In Figure \ref{r2f} we plot the mean square displacements 
as a function of time. 
The numerical results obtained with Montecarlo simulations (symbols) 
are compared with
the behavior predicted by formula (\ref{r2com}), represented by 
thin (red) lines. An excellent agreement is present in the whole range of $\alpha$'s.

In Figure \ref{Prt0_3} we plot the probability density $P(r,t)$ for 
$\alpha=0.3$ at different times. The  scaling hypothesis (\ref{sal}) 
is well satisfied if $\ell(t)$ grows as $t^{1/(1+\alpha)}$. We also verify 
the behaviour for large $r$ predicted by equation (\ref{Plmin1}).
Figure \ref{Prt1_3} evidences that, for $1<\alpha<3/2$, 
$P(r,t)$  presents at large distances a different behaviour. The probability
always decreases as $(r/\ell(t))^{-1-\alpha}$ but the coefficient depends
on time, vanishing as $t^{(1-\alpha)/2}$. In conclusion, the numerical 
simulations confirm the single long jump hypothesis that we have 
introduced in the calculation of $P(r,t)$ both for $0<\alpha<1$ and for
$\alpha \geq 1$.

In this paper, we have presented an analysis of  a one dimensional L\'evy walk
in quenched disorder with correlation induced by the topology. Using the mapping with
the equivalent electric problem and the scaling hypothesis for the random walk probability, 
we have proposed a "single jump" approximation leading to a new estimate for the mean
square displacement and for its moments, when averaged over the scattering points, as a
function of the parameter of the L\'evy distribution for the spacings. Our hypothesis is able to capture
correctly the slow decaying tail effect of the $P(r,t)$, leading to an explicit difference between the scaling
length $\ell^2(t)$ and $\langle r^2(t)\rangle$, which is different from the analogous result originating
from the contribution of the ballistic peak, observed in previous works. Interestingly, in experimental settings with tuned 
L\'evy disorder \cite{Levy}, the average over the scattering points appears to be the relevant quantity to compare
with experimental results. Our findings open the way to an extension of analogous arguments to higher
dimensional disordered samples, where the effects of quenched disorder is still unclear, in view of future experiments.

We acknowledge useful discussion with P. Barthelemy, J. Bertolotti, S. Lepri, R. Livi,
K. Vynck and D.S. Wiersma. This work is partially supported by the MIUR project 
PRIN 2008 \textit{Non linearity and disorder in classical and quantum processes}.

\end{document}